\documentclass[a4paper,12pt]{article}

\usepackage[unicode=true,hypertexnames=false]{hyperref}
\usepackage{amsmath,amssymb,amsthm,amsfonts}
\usepackage{xspace}
\usepackage{xcolor}
\usepackage{cite}
\usepackage{algorithm,algpseudocode}
\usepackage{marginnote}

\usepackage{etoolbox}
\usepackage{tikz}
\usetikzlibrary{arrows.meta,calc,positioning}
\usetikzlibrary{external}
\makeatletter
\let\pgfmathModX=\pgfmathMod@
\usepackage{pgfplots}%
\let\pgfmathMod@=\pgfmathModX
\makeatother

\graphicspath{{pic/}}

\pgfplotsset{compat=1.14}
\pgfdeclarelayer{background}
\pgfdeclarelayer{foreground}
\pgfsetlayers{background,main,foreground}

\newcommand{\TikZPictureSweepAShot}[9]{
    \draw[->] (0,0)--(10,0);
    \draw[->] (0,0)--(0,8);
    \draw[dashed] (#1,0)--(#1,7.5);

    \node[passive] (a1) at (1, 5) {1};
    \node[passive] (b1) at (2, 2) {1};
    \node[active]  (c1) at (5, 1) {1};

    \node[passive] (a2) at (3, 4) {2};
    \node[#8]      (b2) at (5, 3) {2};

    \node[passive] (a3) at (4, 6) {3};
    \node[#8]      (b3) at (6, 4) {3};
    \node[#2]      (c3) at (7, 3) {#3};

    \node[#9]      (a4) at (5, 7) {4};
    \node[#4]      (a5) at (8, 2) {5};
    \node[#5]      (a6) at (9, 6) {#6};

    \draw (a1)--(b1)--(c1);
    \draw (a2)--(b2);
    \draw (a3)--(b3)#7;
}

\newcommand{\TikZPictureSweepA}{
    \tikzset{default/.style={draw,circle,inner sep=1.5pt}}
    \tikzset{passive/.style={default,fill=gray!30}}
    \tikzset{active/.style={default,fill=gray!70}}
    \tikzset{queued/.style={default,fill=white}}

    \begin{scope}[xshift=0cm, yshift=0cm]
        \TikZPictureSweepAShot{7}{queued}{1}{queued}{queued}{2}{}{active}{active}
        \draw[dotted] (c1)--(b2)--(b3)--(a4);
        \node (label) at (1, 8) {(a)};
    \end{scope}
    \draw[->] (10,4)--(11,4);
    \begin{scope}[xshift=12cm, yshift=0cm]
        \TikZPictureSweepAShot{8}{active}{3}{queued}{queued}{2}{--(c3)}{passive}{active}
        \draw[dotted] (c1)--(c3)--(a4);
        \node (label) at (1, 8) {(b)};
    \end{scope}
    \draw[->] (22,4)--(23,4);
    \begin{scope}[xshift=24cm, yshift=0cm]
        \TikZPictureSweepAShot{9}{passive}{3}{active}{queued}{2}{--(c3)}{passive}{passive}
        \draw[dotted] (c1)--(a5);
        \node (label) at (1, 8) {(c)};
    \end{scope}
    \draw[->] (34,4)--(35,4);
    \begin{scope}[xshift=36cm, yshift=0cm]
        \TikZPictureSweepAShot{10}{passive}{3}{active}{active}{6}{--(c3)}{passive}{passive}
        \draw[dotted] (c1)--(a5)--(a6);
        \node (label) at (1, 8) {(d)};
    \end{scope}
}

\input{pic/plots.tex}

\begin{document}

\title{Hybridizing Non-dominated Sorting Algorithms: Divide-and-Conquer Meets\\Best Order Sort}

\author{Margarita Markina \and Maxim Buzdalov}

\maketitle

\begin{abstract}
Many production-grade algorithms benefit from combining an asymptotically efficient algorithm for solving big problem instances, by splitting them into smaller ones,
and an asymptotically inefficient algorithm with a very small implementation constant for solving small subproblems. A well-known example is stable sorting,
where mergesort is often combined with insertion sort to achieve a constant but noticeable speed-up.

We apply this idea to non-dominated sorting. Namely, we combine the divide-and-conquer algorithm,
which has the currently best known asymptotic runtime of $O(N (\log N)^{M - 1})$,
with the Best Order Sort algorithm, which has the runtime of $O(N^2 M)$
but demonstrates the best practical performance out of quadratic algorithms.

Empirical evaluation shows that the hybrid's running time is typically not worse than of both original algorithms,
while for large numbers of points it outperforms them by at least 20\%. For smaller numbers of objectives,
the speedup can be as large as four times.
\end{abstract}

\section{Introduction}

Many real-world optimization problems are multiobjective, that is, they require maximizing or
minimizing several objectives, which are often conflicting. These problems most often do not have
a single solution, but instead feature many incomparable solutions, which trade one objective for another.
It is often not known \emph{a priori} which solution will be chosen, as decisions of this sort
are often recommended to be made late, as the decision maker can learn more about the problem~\cite{gecco-2016-emo-tutorial}.
This encourages finding a set of diverse incomparable solutions,
which is a problem often approached by multiobjective evolutionary algorithms.

In the realm of scaling-independent preference-less, and thus general-purpose, evolutionary multiobjective algorithms,
three paradigms currently seem to prevail~\cite{gecco-2016-emo-tutorial}:
Pareto-based, indicator-based, and decomposition-based approaches.
Although there exist well-known decomposition-based~\cite{moea-d} and indicator-based~\cite{hypervolume,ibea,epsilon-indicator} algorithms,
the majority of modern algorithms are Pareto-based~\cite{nsga-ii,nsga-iii,pesa-ii,spea2}.

Most Pareto-based algorithms belong to one of big groups according to how solutions are selected or ranked:
the algorithms which maintain non-dominated solutions~\cite{pesa-ii,paes,micro-ga},
perform non-dominated sorting~\cite{nsga-ii,nsga-iii,npga2}, use domination count~\cite{fonseca-moga}, or domination strength~\cite{spea2}.
In this research we concentrate on non-dominated sorting, as some popular algorithms make use of it~\cite{nsga-ii,nsga-iii}.

Non-dominated sorting assigns ranks to solutions in the following way: the non-dominated solutions get rank 0,
and the solutions which are dominated only by solutions of rank at most $i$ get rank $i+1$. In the original work~\cite{nsga},
this procedure was performed in $O(N^3 M)$, where $N$ is the population size and $M$ is the number of objectives. This was later improved
to be $O(N^2 M)$ in~\cite{nsga-ii}.

As the quadratic complexity is still quite large, both from theoretical and practical points of view,
many researchers concentrated on improving practical running times~\cite{zhang-ens,zhang-ens-presort,fang-nds-tree,gustavsson-nds,best-order-sort-gecco,deductive-climbing-sort,corner-sort},
however, without improving the worst-case $O(N^2 M)$ complexity. Jensen was the first to adapt the earlier result of Kung et at.~\cite{kung75},
who solved the problem of finding non-dominated solutions in $O(N (\log N)^{\max(1, M-2)})$, to non-dominated sorting.
This algorithm has the worst-case complexity of $O(N (\log N)^{M-1})$. However, this algorithm could not handle coinciding objective values,
which was later corrected in subsequent works~\cite{fortin,buzdalov-nds-2014}. A more efficient algorithm for non-dominated sorting,
or finding \emph{layers of maxima}, exists for three dimensions~\cite{nekrich-fast-3d-nds}, whose complexity is
$O(N (\log \log N)^2)$ with the use of randomized data structures, or $O(N (\log \log N)^3)$ for deterministic ones.
However, whether this algorithm is useful in practice is still an open question.

A large number of available algorithms for non-dominated sorting opens the question of algorithm selection~\cite{rice-algorithm-selection}.
What is more, a family of $O(N^2 M)$ algorithms for non-dominated sorting resembles a family of quadratic algorithms for comparison based sorting,
and the $O(N (\log N)^{M-1})$ non-dominated sorting algorithms seem to take up the niche of $O(N \log N)$ sorting algorithms (such as mergesort,
heapsort, and randomized versions of quicksort).

For comparison-based sorting, the quadratic algorithms are often much simpler and demonstrate better performance on small data,
while asymptotically better algorithms take over starting from certain problem sizes. If the latter algorithm is built using a divide-and-conquer
scheme, it becomes possible to choose better algorithms for subproblems: if a subproblem, due to its size, can be solved faster using a quadratic
algorithm, then it should be done, otherwise let the divide-and-conquer algorithm decompose the problem further.
For example, most stable sorting algorithms from standard libraries are currently implemented using mergesort or TimSort,
while for data fragments smaller than, for example, 32 in the current implementation of sorting in 
Java\footnote{\url{http://grepcode.com/file_/repository.grepcode.com/java/root/jdk/openjdk/8-b132/java/util/TimSort.java}},
the quadratic insertion sort algorithm, with the binary search lookup, is used.

This inspired us to apply the similar idea to non-dominated sorting. For the ``outer'' divide-and-conquer algorithm,
we use the only available algorithm family of this sort~\cite{jensen,fortin,buzdalov-nds-2014}. For the quadratic algorithm
to solve smaller subproblems, we adapt the Best Order Sort~\cite{best-order-sort-gecco}, as it was shown to typically outperform
other quadratic algorithms. Our result is a hybrid algorithm which uses primarily the divide-and-conquer strategy
and decides when to switch to Best Order Sort using a formula which depends on the number of points in the subproblem
and the number of remaining objectives to consider.

This is a full version of the paper with the same name which was accepted as a poster to the GECCO conference in 2017.

The rest of the paper is structured as follows. In Section~\ref{preliminaries}, we give the necessary definitions
and describe the algorithms we put together: the divide-and-conquer algorithm and Best Order Sort.
Section~\ref{hybridizing} describes our hybridizing approach, which includes the changes necessary to introduce to Best Order Sort
to serve as the subproblem solver, and the analysis of preliminary experiments which established the formula used to switch between the algorithms.
Section~\ref{experiments} gives the main body of our experimental studies, including their analysis.
Section~\ref{conclusion} concludes.

\section{Preliminaries}\label{preliminaries}

In the following, we assume that all points are different, which enables us to name any unordered collection of points a \emph{set}.
This is not true in general, however, all equal points will receive the same rank, so implementations are free, depending on their need,
to either discard a point if there is an equal one, or to keep all equal points in a same entity and run algorithms on these entities instead,
or to work directly with equal points with some additional algorithmic care. None of these precautions change the worst-case algorithmic complexity.

\subsection{Definitions}

We use capital Latin letters to denote sets of points, as well as the global constants $N$ (the
number of points) and $M$ (the number of objectives),
while small Latin letters are used for single points, standalone objectives and rank values,
and small Greek letters are used for mappings. The value of the $i$-th objective of a point $p$ is
denoted as $p_i$.

In the rest of the paper we assume, without losing generality, that we solve a multiobjective minimization problem
with the number of objectives equal to $M$.
In this case, the \emph{Pareto dominance relation} is determined on two points in the objective space as follows:
\begin{align*}
    a \prec b   &\leftrightarrow \forall i \in [1; M] \; a_i \le b_i
                    \text{ and } \exists i \in [1; M] \; a_i < b_i \\
    a \preceq b &\leftrightarrow \forall i \in [1; M] \; a_i \le b_i
\end{align*}
where $a \prec b$ is called the \emph{strict dominance} and $a \preceq b$ is the \emph{weak dominance}.

\emph{Non-dominated sorting} is a procedure which, for a given set $P$ of $N$ points in the $M$-dimensional objective space,
assigns each point $p \in P$ an integer \emph{rank} $\tau(p)$, such that:
\begin{align*}
\tau(p) = \max \{0\} \cup \{ 1 + \tau(q) \mid q \in P, q \prec p \}.
\end{align*}
In other words, a rank of a point which is not dominated by any other point is zero,
and a rank of any other point is one plus the maximum rank among the points which dominate it.

Following the convention from~\cite{deb-enlu-14},
we call the set of all points with the given rank $r$ a \emph{non-domination level} $L_r$:
\begin{align*}
L_r = \{ p \in P \mid \tau(p) = r \}.
\end{align*}

\subsection{The Divide-and-Conquer Approach}

The divide-and-conquer approach dates back to 1975, when Kung et al. proposed a multidimensional divide-and-conquer algorithm
for finding the maxima of a set of vectors~\cite{kung75}, which, in the realm of evolutionary computation,
corresponds to the set of non-dominated points, or to points with rank zero.
The complexity of this algorithm is $O(\min(N^2, N (\log N)^{\max(1, M - 2)})))$,
which we shorten to $O(N (\log N)^{M - 2})$ for clarity.

This algorithm can be used to implement non-dominated sorting in the following manner: first we determine the points with rank zero,
then we remove these points and run the algorithm again on the remaining points (which yields points with rank one),
then we repeat it until no points left. However, the worst-case complexity of this approach is $O(N^2 (\log N)^{M - 2})$.
In contrast, fast non-dominated sorting, shipped with the original NSGA-II of Deb et at.~\cite{nsga-ii}, has a better $O(N^2 M)$ complexity.

The divide-and-conquer approach has been generalized to perform non-dominated sorting by Jensen~\cite{jensen},
shortly afterwards the NSGA-II arrived. The algorithm from~\cite{jensen} solves the problem in $O(N (\log N)^{M - 1})$,
which is much faster for small values of $M$, as well as for large values of $N$, than fast non-dominated sorting.
However, this algorithm was designed with an assumption that no two points have equal objectives, which is often not the case,
especially in discrete optimization, and is known to produce wrong results when this assumption is violated.
This problem was overcome by Fortin et al.~\cite{fortin}, who proposed modifications of this algorithm to always produce
correct results. The average complexity was proven to be the same, but the worst-case complexity was left at $O(N^2 M)$.
Finally, Buzdalov et al.~\cite{buzdalov-nds-2014} introduced further modifications to achieve the worst-case
time complexity of $O(N (\log N)^{M - 1})$.

We shall now briefly illustrate the working principles of this approach. At any moment of time,
the algorithm maintains, for every point $p$, a lower bound on its rank $r'(p)$, which are initially
set to zero. The reason for this lower bound can be explained as follows: at any moment of time,
we have performed a subset of necessary objective comparisons, which impose approximations of ranks
of the affected points. These approximations are of course lower bounds of the real ranks.

To ease the notation, in the following we do not use the term ``lower bound of the rank'', as well as the $r'$ symbol.
Instead, we will say ``current rank'' for the current state of the lower bound of a certain point, which possibly
coincides with the real rank, and ``final rank'' when we know that the lower bound coincides with the real rank.

One of the main properties of the algorithm is that whenever a comparison of $p_m$ and $q_m$ is performed for the first time,
where $p$ and $q$ are points and $1 \le m \le M$ is the objective, then the following holds:
\begin{itemize}
    \item for all objectives $m'$ such that $m < m' \le M$, it holds that $p_{m'} \le q_{m'}$, that is,
          $p$ weakly dominates $q$ in objectives $[m';M]$;
    \item the rank of $p$ is known and final, that is, all comparisons necessary to determine the rank of $p$
          have already been done.
\end{itemize}

The top-level concept is the procedure $\textsc{HelperA}(S, m)$, which takes a set of points $S$ sorted lexicographically
(where non-zero lower bounds are possibly known for some of the points from $S$)
and makes sure all necessary comparisons between the objectives $[1;m]$ of these points are performed.
This procedure is called only when all necessary comparisons of points $p$ and $q$, such that $q \in S$ and $p \notin S$,
have already been performed.
To perform non-dominated sorting of a set $P$ with $M$ objectives, one should run $\textsc{HelperA}(P, M)$.

For $m = 2$, it calls a sweep line based algorithm $\textsc{SweepA}(S)$, which runs in $O(|S| \log |S|)$, which we will cover later.
If there are at most two points in $S$, it performs their direct comparisons and updates the rank of the
second point if necessary. If all values of the objective $m$ are the same in the entire $S$,
it directly calls $\textsc{HelperA}(S, m - 1)$.
Otherwise, it divides $S$ into three parts using the objective $m$:
the $S_L$ part with lower values, the $S_M$ part with median values,
and the $S_H$ part with higher values.

It is clear that ranks of points in $S_L$ do not depend on ranks of points in neither $S_M$ nor $S_H$,
and $S_M$ also does not depend on $S_H$. The algorithm first calls $\textsc{HelperA}(S_L, m)$,
which results in finding the exact ranks in $S_L$, because all necessary comparisons with points from $S_L$ on the right side
and other points on the left side have been performed before this call.

Next comes the set $S_M$, but the ranks of these points still need to be updated using the set $S_L$ (and nothing more).
To do this, the algorithm calls another procedure, $\textsc{HelperB}(S_L, S_M, m - 1)$, whose meaning is to update
the ranks of points from the second argument using the first argument and objectives in $[1; m-1]$.
Then it calls $\textsc{HelperA}(S_M, m - 1)$, as all other necessary comparisons have been done, and all values for the objective $m$
are equal in $S_M$. It then proceeds with $\textsc{HelperB}(S_L \cup S_M, S_H, m - 1)$ and finishes with $\textsc{HelperA}(S_H, m)$.

The $\textsc{HelperB}(L, H, m)$ procedure, as follows from the short description above, shall perform all the necessary comparisons
between points $p \in L$ on the left and $q \in H$ on the right, provided that in objectives $[m+1; M]$ it holds that $p \prec q$,
and all ranks in $L$ are final. For $m = 2$, it, again, runs a sweep line procedure $\textsc{SweepB}(L, H)$.
If $|L| = 1$ or $|H| = 1$, a straightforward pairwise comparison is performed. If the maximum value of the objective $m$ in $L$
does not exceed the minimum value in $H$, it calls $\textsc{HelperB}(L, H, m-1)$. Otherwise, it chooses a median of the objective $m$
in $L \cup H$ and then, similarly to \textsc{HelperA}, splits $L$ into $L_L$, $L_M$ and $L_H$,
and also splits $H$ into $H_L$, $H_M$ and $H_H$. Following the same logic as in \textsc{HelperA}, it performs the following
recursive calls:
\begin{itemize}
    \item $\textsc{HelperB}(L_L, H_L, m)$;
    \item $\textsc{HelperB}(L_L, H_M, m - 1)$;
    \item $\textsc{HelperB}(L_M, H_M, m - 1)$;
    \item $\textsc{HelperB}(L_L \cup L_M, H_H, m - 1)$;
    \item $\textsc{HelperB}(L_H, H_H, m)$.
\end{itemize}

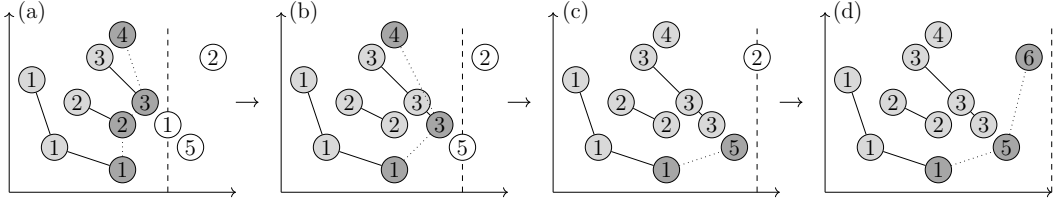
\begin{figure*}[!t]
\scalebox{0.71}{
\begin{tikzpicture}[scale=0.42]
\TikZPictureSweepA
\end{tikzpicture}
}
\caption{Example iterations of the \textsc{SweepA} procedure.
Gray points are those whose rank is determined, the darker ones
constitute a binary search tree and thus connected with dotted lines.
The numbers in white points are the lower bounds on ranks.
The vertical dashed line is the sweep line.
In (b), the representer of level 2 is removed as all possible remaining
points dominated by that point are also dominated by the representer of level 3;
similar thing happens in (c).
}\label{sweep-a-example}
\end{figure*}

The remaining parts to explain are $\textsc{SweepA}(S)$ and $\textsc{SweepB}(L, H)$.
The \textsc{SweepA} procedure utilizes a sweep line approach. Points from the set $S$
are processed in lexicographical order using first two objectives. In the same time,
the procedure maintains a binary search tree which contains the last seen representative points
for each non-domination level. When the next point is processed, this binary search tree is traversed
to determine the biggest number of the level which still dominates the point in question, and then
the rank of this point is updated correspondingly. After that, this point is inserted in the tree:
it becomes the last representative of its non-domination level and possibly throws out some of
the other representatives, which have no more chance to determine rank of any point on their own.
An example is shown in Fig.~\ref{sweep-a-example}.

The \textsc{SweepB} procedure works in a similar way. The sweep line goes over the union of sets, $L \cup H$,
however, the tree is built of the points from $L$ only, and rank updates are performed with points from $H$ only.

The running times of \textsc{SweepA} and \textsc{SweepB} are $O(|S| \log |S|)$ and $O((|L| + |H|) \log |L|)$,
respectively. From the well-known theory of solving recursive relations, and from the strategies of creating subproblems,
it follows that the running time of $\textsc{HelperB}(L, H, m)$ is $O((|L| + |H|) \cdot (\log (|L| + |H|))^{m - 1})$,
and of $\textsc{HelperA}(S, m)$ it is $O(|S| \cdot (\log |S|)^{m - 1})$.

\subsection{Best Order Sort}

The Best Order Sort algorithm was proposed in~\cite{best-order-sort-gecco}.
It aims at removing as many comparisons to be performed as possible.
To do this, it sorts all points by all objectives, thus constructing $M$ sorted
lists of points $L_1 \ldots L_M$, and processes the points in the following
order: first, all first points in the lists ($L_{1,1}, \ldots, L_{M,1}$),
then all second points ($L_{1,2}, \ldots, L_{M,2}$), then all third points,
et cetera, until every point is processed at least once.

When a point $p$ is processed for the first time, assume it happens in the list of the $m$-th objective,
its rank has to be determined. The key fact is that only the points which precede $p$ in $L_m$ can dominate
$p$, because all other points have a greater value of the $m$-th objective. Thus, it makes sense to compare
$p$ with the points that precede it in $L_m$.

To further decrease the number of comparisons, it is worth noting that,
when a certain point $p$ is processed in objective $m$, all subsequent \emph{new} points, that is the points
which will be processed for the first time, will have a value of the $m$-th objective which is not smaller
than the one of $p$. This means that the objective $m$ can be safely removed from the list of objectives
to test when some other point $q$ is checked for being dominated by $p$.

The algorithm maintains a set of objectives to consider $O_p$ for every point $p$. Initially, 
$O_p \gets \{ 1, 2, \ldots, M\}$. Whenever a point $p$ is processed in the list of the $m$-th objective, 
it is removed from $O_p$. Whenever a point $q$ is checked for being dominated by $p$,
only the objectives from $O_p$ need to be considered.

Finally, to determine the rank using fewer comparisons, the points, which have been already considered
in each objective list and have been assigned ranks, are stored in separate lists, where each list
corresponds to a rank. To determine the rank of the next point, one can perform either a linear scan
(starting with rank zero and increasing ranks by one) or binary search for the rank.
As the number of points in rank lists cannot be non-trivially bounded, both ways have the worst-case
complexity of a single search of $O(LM)$, where $L$ is the number of points in all lists.

Best Order Sort features two phases: the pre-sorting phase, which takes $O(NM \log N)$, and the
domination scanning phase. The complexity of the latter, in the worst case, is $O(N^2 M)$,
but can be smaller under various conditions. For instance, when all points are non-dominating,
the points have a chance to arrange such that the first $N$ processed points are unique, which means that
every such point is tested against $O(N / M)$ points in average, which results in $O(N^2)$ running time.


\section{Hybridizing the Algorithms}\label{hybridizing}

Our hybridization scheme is similar to that of production-grade sorting algorithms tuned for performance.
As the top-level algorithm, we use the divide-and-conquer algorithm. For each subproblem it decides,
using certain heuristic, whether to continue using the divide-and-conquer strategy or to
run Best Order Sort for this subproblem.
In turn, Best Order Sort runs uninterrupted until it solves the assigned subproblem.

Two problems need to be solved for this scheme to work. First, the original Best Order Sort algorithm
cannot be straightforwardly applied to solve subproblems, because subproblems may feature non-zero lower
bounds for ranks of some points, which appear from comparisons of these points with other points, which are
out of the scope of the current subproblem. It also does not support working with two point sets
in order to serve as a back-end of \textsc{HelperB}.

Second, the particular kind of heuristic to determine when to run Best Order Sort is unclear. The main problem
with it is that it should have a low computation complexity: at most $O(N)$, because otherwise evaluation of this
heuristic worsens the complexity of the divide-and-conquer algorithm. This means we cannot perform any complicated
analysis, such as, for instance, principal component analysis, to predict which algorithm is best.

In this section we address these two problems, which determines the shape of our hybridization approach.

\subsection{Adaptation of Best Order Sort}

When working as a part of the divide-and-conquer algorithm, Best Order Sort can be called instead
either \textsc{HelperA} or \textsc{HelperB}. In the first case, it needs to assign final ranks
to a set of points $S$ using first $m$ objectives ($m > 2$, as \textsc{SweepA}, due to its simplicity,
works faster than Best Order Sort under any conditions), provided that all other necessary comparisons
have been already performed, and consequently every point $p$ has a current rank $\tau(p)$, which is a lower
bound of its real rank. The only difference to the original Best Order Sort is that some $\tau(p)$ can be non-zero.
This is easily compensated by checking only rank lists with ranks greater than or equal to $\tau(p)$,
and thus updating the rank only if the update is increasing.

The \textsc{HelperB} case is slightly more involved. If Best Order Sort is called within $\textsc{HelperB}(L, H, m)$,
then ranks of points from $L$ are already known, and it is necessary to perform comparisons between points from $L$ and points from $H$
to update the current ranks $\tau(p)$ of points $p \in H$ using first $m$ objectives.
In this case, all points are merged and are processed altogether. However, for points from $L$ the rank is not updated
(that is, the rank lists are never checked), instead they go directly to the corresponding rank lists.
On the contrary, the rank update procedure is executed on points from $H$, but they are never added to rank lists.

These changes are quite small, so the correctness of Best Order Search in the changed conditions
follows straightforwardly from the correctness of the original algorithm~\cite{best-order-sort-gecco}.
The worst-case complexity of the \textsc{HelperB} case is $O(M \cdot |L \cup H| \cdot \log |L \cup H|, M \cdot |L| \cdot |H|)$.

\subsection{Design of the Switch Heuristic}

\begin{figure}[!t]
\centering\includegraphics[width=0.7\columnwidth]{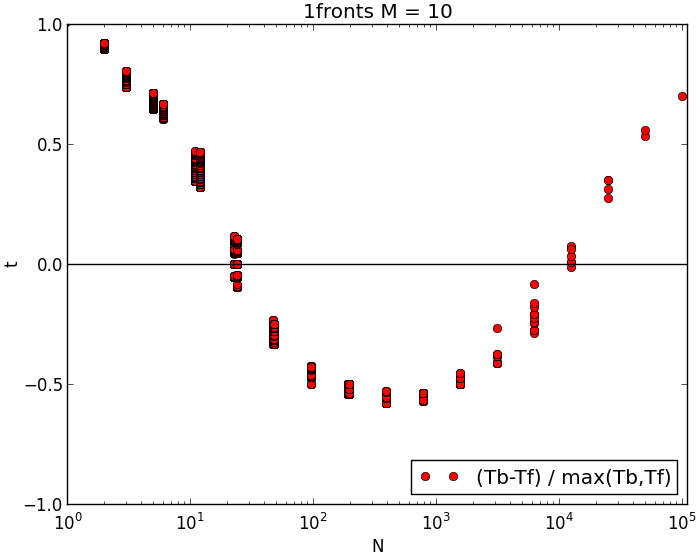}
\caption{Example result of preliminary experiments on a dataset with 10 objectives and one non-domination level.
The dataset has $N = 10^5$ points, all other points correspond to divide-and-conquer subproblems
for this dataset. $T_f$ is the running time of the divide-and-conquer algorithm,
and $T_b$ is the running time of Best Order Sort. The value of $(T_b - T_f) / \max(T_f, T_b)$ is plotted.
}\label{example-plot-dim10-front1}
\end{figure}

To understand the possible kind of the heuristic algorithm to use for deciding whether to use Best Order Sort for a certain subproblem,
we conducted a series of preliminary experiments. In these experiments, we considered a series of datasets,
where every dataset had $N = 10^5$ points with $M \in [3; 20]$ objectives and was generated either by uniformly random objective sampling
(from the $[0;1]^M$ hypercube) or by sampling from a hyperplane (which yields a dataset with exactly one non-domination level).
Then we ran the divide-and-conquer algorithm on each of these datasets and recorded all subproblems created during the run.
After that, we measured the running times of both the divide-and-conquer algorithm and Best Order Sort on all these subproblems.

Fig.~\ref{example-plot-dim10-front1} shows an example of such experiment. In this figure, the point above the abscissa axis
means that for the corresponding subproblem the divide-and-conquer algorithm took less time than Best Order Sort,
while a point below zero means the opposite. One can clearly see in Fig.~\ref{example-plot-dim10-front1} that Best Order Sort
behaves best, compared to the divide-and-conquer algorithm, for $N$ which are not too small and not too large.

\begin{figure}[!t]
\centering\includegraphics[width=0.7\columnwidth]{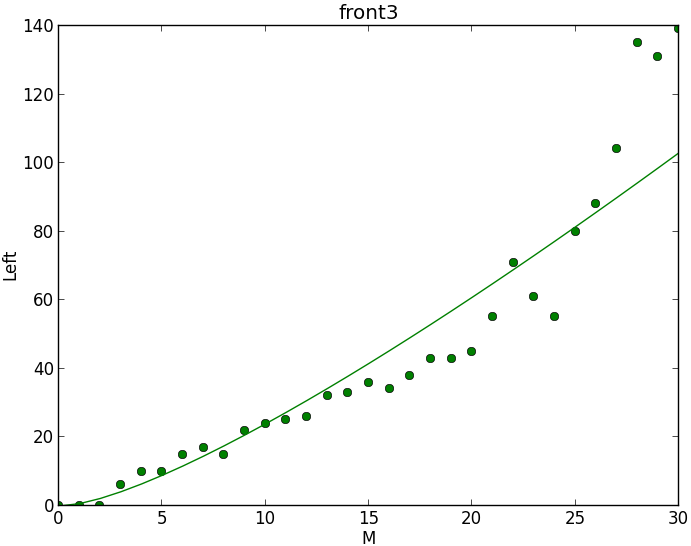}
\caption{Left bounds of the BOS-efficient range: actual bounds from datasets with three levels and the fitted curve}\label{example-left}
\end{figure}

\begin{figure}[!t]
\centering\includegraphics[width=0.7\columnwidth]{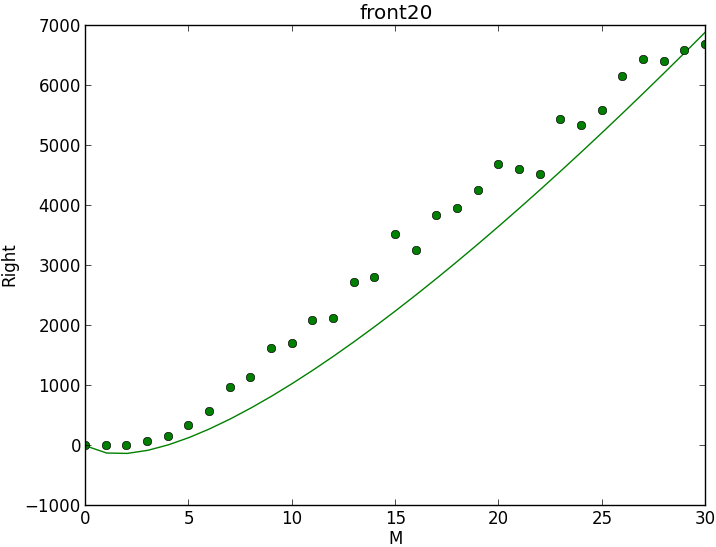}
\caption{Right bounds of the BOS-efficient range: actual bounds from datasets with twenty levels and the fitted curve}\label{example-right}
\end{figure}

As the similar effect has been noticed for all other datasets as well, we attempted to deduce formulas for the left and right bounds
of the higher efficiency range of Best Order Sort. The following empirically constructed formulas were found to fit our data rather well: 
$n_{\min} = m \ln (m + 1)$ and $n_{\max} = 150 m ((\ln (d + 1))^{0.9} - 1.5)$, where $m$ is the current number of first objectives to consider,
$n_{\min}$ is the left bound of the range, and $n_{\max}$ is the right bound.
Fig.~\ref{example-left} shows the plot of the left bound formula and the actual left bounds in datasets with three non-domination levels,
Fig.~\ref{example-right} does the same for the right bound formula and datasets with twenty levels.
The fitting quality is the same for all other considered datasets.

As a result, the hybrid algorithm switches to Best Order Sort whenever the number of points $n$ and the number of considered objectives $m$
in the current subproblem satisfy: \[m \ln (m + 1) \le n \le 150 m \cdot ((\ln (d + 1))^{0.9} - 1.5).\]

\section{Experiments}\label{experiments}

The main part of experiments was organized as follows. For every combination of:
\begin{itemize}
    \item numbers of points $N = \lfloor 10^{n / 4} \rfloor$ where $n \in [8; 20]$;
    \item numbers of objectives $M \in \{ 3, 5, 7, 10, 15, 20, 25, 30 \}$;
    \item numbers of non-domination levels $L \in \{ 1, 2, 3, 5, 10, 20 \}$;
\end{itemize}
ten datasets were created, and running times of all considered algorithms (the divide-and-conquer algorithm, Best Order Sort,
and the hybrid algorithm) were measured.

The results are presented on pages~\pageref{first-fig}--\pageref{last-fig} as bar plots for all $M$ and $L$. Each bar plot features a section corresponding to the
value of $N$, consisting of the following three bars: $T_{BOS} / \text{avg}(T_{DC})$, $T_{DC} / \text{avg}(T_{DC})$, $T_H / \text{avg}(T_{DC})$,
where $T_{BOS}$ is the running time of Best Order Sort, $T_{DC}$ is the running time of the divide-and-conquer algorithm,
and $T_H$ is the running time of the proposed hybrid algorithm.
The bars for Best Order Sort are blue, and the bars for the hybrid algorithm are brown.
Every bar has an average, minimum and maximum value (for the second
bar plotting $T_{DC}$, the average is always one). Whenever a bar's average is greater than one, that is, it points up,
it means that the corresponding algorithm is slower than divide-and-conquer, and if it is faster, then the bar points down.

From plots on pages~\pageref{first-fig}--\pageref{last-fig}, one can immediately spot the characteristic behavior of Best Order Sort: it is typically better 
at smaller numbers of points, then it gradually becomes worse (for $M = 7$ and $M = 10$, this tendency is seen the best).
For somwhat higher dimensions ($M = 20$ and $M = 25$), the lower bound of the Best Order Sort efficiency interval can be seen.
For the highest considered dimension, $M = 30$, Best Order Sort demonstrates no significant improvement over the divide-and-conquer algorithm.

The hybrid algorithm tends to perform at least as good as the best of the two algorithms up to $M = 7$. Starting from $M = 10$, it features
a somewhat suboptimal performance at the middle problem sizes $(N \in [10^3; 10^4])$ while still capturing the best behavior at small sizes
and getting better than all other algorithms close to $N = 10^5$.

In fact, the hybrid is always better than its parts for big numbers of points. For $N = 10^5$, the average speedup compared to the best of the parts
can be as large as $4.28$ when $M = 3$, and never seen to get less than $1.198$ in all other considered datasets.

\section{Conclusion}\label{conclusion}

We presented a hybrid algorithm for non-dominated sorting which initially runs a divide-and-conquer algorithm,
however, when the size of a certain subproblem seems to be suitable, it solves this subproblem using another approach,
Best Order Sort. For this to work, we slightly adapted Best Order Sort, so that it can perform non-dominated sorting
in a more general setup, which needs to solve the divide-and-conquer subproblems. We also composed a heuristic rule
for when to switch to Best Order Sort, which is based solely on the dimensions of a subproblem.

Our algorithm performs generally at least as well as its parts, except for certain ranges around the
switchpoint between the algorithms at higher dimensions. This is an indicator that our heuristic on when to switch is not perfect yet
and has a room for improvement. Nevertheless, for the wide range of testing data (3 to 30 objectives, 1 to 20 non-domination levels)
our algorithm performs at least 20\% better than the best of its parts for large numbers of points (such as $N = 10^5$),
and the speedup can be up to 4x for smaller $M$. In a sense, this means that our hybridization scheme is rather robust. 

\bibliographystyle{abbrv}
\bibliography{../../../../bibliography}

\newcommand{\includethreewide}[4]{
\begin{tabular}{c}
\begin{tikzpicture}[scale=0.72]
\begin{axis}[ybar, bar width=1pt, xmode=log, ymode=log, xtick pos=left]
\addplot plot[error bars/.cd, y dir=both, y explicit] table[y error plus=y-max, y error minus=y-min] {#2};
\addplot plot[error bars/.cd, y dir=both, y explicit] table[y error plus=y-max, y error minus=y-min] {#1};
\addplot plot[error bars/.cd, y dir=both, y explicit] table[y error plus=y-max, y error minus=y-min] {#3};
\end{axis}
\end{tikzpicture} \\
#4
\end{tabular}}

\newpage

\begin{figure}[!t]
\begin{tabular}{rr}
\includethreewide{\FastDthreeFone}{\BOSDthreeFone}{\HybridDthreeFone}{$M = 3$, one level} & 
\includethreewide{\FastDthreeFtwo}{\BOSDthreeFtwo}{\HybridDthreeFtwo}{$M = 3$, two levels} \\
\includethreewide{\FastDthreeFthree}{\BOSDthreeFthree}{\HybridDthreeFthree}{$M = 3$, three levels} &
\includethreewide{\FastDthreeFfive}{\BOSDthreeFfive}{\HybridDthreeFfive}{$M = 3$, five levels} \\
\includethreewide{\FastDthreeFten}{\BOSDthreeFten}{\HybridDthreeFten}{$M = 3$, 10 levels} &
\includethreewide{\FastDthreeFtwenty}{\BOSDthreeFtwenty}{\HybridDthreeFtwenty}{$M = 3$, 20 levels}
\end{tabular}
\label{first-fig}
\end{figure}

\begin{figure}[!t]
\begin{tabular}{rr}
\includethreewide{\FastDfiveFone}{\BOSDfiveFone}{\HybridDfiveFone}{$M = 5$, one level} & 
\includethreewide{\FastDfiveFtwo}{\BOSDfiveFtwo}{\HybridDfiveFtwo}{$M = 5$, two levels} \\
\includethreewide{\FastDfiveFthree}{\BOSDfiveFthree}{\HybridDfiveFthree}{$M = 5$, three levels} &
\includethreewide{\FastDfiveFfive}{\BOSDfiveFfive}{\HybridDfiveFfive}{$M = 5$, five levels} \\
\includethreewide{\FastDfiveFten}{\BOSDfiveFten}{\HybridDfiveFten}{$M = 5$, 10 levels} & 
\includethreewide{\FastDfiveFtwenty}{\BOSDfiveFtwenty}{\HybridDfiveFtwenty}{$M = 5$, 20 levels}
\end{tabular}
\end{figure}

\begin{figure}[!t]
\begin{tabular}{rr}
\includethreewide{\FastDsevenFone}{\BOSDsevenFone}{\HybridDsevenFone}{$M = 7$, one level} &
\includethreewide{\FastDsevenFtwo}{\BOSDsevenFtwo}{\HybridDsevenFtwo}{$M = 7$, two levels} \\
\includethreewide{\FastDsevenFthree}{\BOSDsevenFthree}{\HybridDsevenFthree}{$M = 7$, three levels} &
\includethreewide{\FastDsevenFfive}{\BOSDsevenFfive}{\HybridDsevenFfive}{$M = 7$, five levels} \\
\includethreewide{\FastDsevenFten}{\BOSDsevenFten}{\HybridDsevenFten}{$M = 7$, 10 levels} & 
\includethreewide{\FastDsevenFtwenty}{\BOSDsevenFtwenty}{\HybridDsevenFtwenty}{$M = 7$, 20 levels}
\end{tabular}
\end{figure}

\begin{figure}[!t]
\begin{tabular}{rr}
\includethreewide{\FastDtenFone}{\BOSDtenFone}{\HybridDtenFone}{$M = 10$, one level} &
\includethreewide{\FastDtenFtwo}{\BOSDtenFtwo}{\HybridDtenFtwo}{$M = 10$, two levels} \\
\includethreewide{\FastDtenFthree}{\BOSDtenFthree}{\HybridDtenFthree}{$M = 10$, three levels} &
\includethreewide{\FastDtenFfive}{\BOSDtenFfive}{\HybridDtenFfive}{$M = 10$, five levels} \\
\includethreewide{\FastDtenFten}{\BOSDtenFten}{\HybridDtenFten}{$M = 10$, 10 levels} & 
\includethreewide{\FastDtenFtwenty}{\BOSDtenFtwenty}{\HybridDtenFtwenty}{$M = 10$, 20 levels}
\end{tabular}
\end{figure}

\begin{figure}[!t]
\begin{tabular}{rr}
\includethreewide{\FastDfifteenFone}{\BOSDfifteenFone}{\HybridDfifteenFone}{$M = 15$, one level} &
\includethreewide{\FastDfifteenFtwo}{\BOSDfifteenFtwo}{\HybridDfifteenFtwo}{$M = 15$, two levels} \\
\includethreewide{\FastDfifteenFthree}{\BOSDfifteenFthree}{\HybridDfifteenFthree}{$M = 15$, three levels} &
\includethreewide{\FastDfifteenFfive}{\BOSDfifteenFfive}{\HybridDfifteenFfive}{$M = 15$, five levels} \\
\includethreewide{\FastDfifteenFten}{\BOSDfifteenFten}{\HybridDfifteenFten}{$M = 15$, 10 levels} &
\includethreewide{\FastDfifteenFtwenty}{\BOSDfifteenFtwenty}{\HybridDfifteenFtwenty}{$M = 15$, 20 levels}
\end{tabular}
\end{figure}

\begin{figure}[!t]
\begin{tabular}{rr}
\includethreewide{\FastDtwentyFone}{\BOSDtwentyFone}{\HybridDtwentyFone}{$M = 20$, one level} &
\includethreewide{\FastDtwentyFtwo}{\BOSDtwentyFtwo}{\HybridDtwentyFtwo}{$M = 20$, two levels} \\
\includethreewide{\FastDtwentyFthree}{\BOSDtwentyFthree}{\HybridDtwentyFthree}{$M = 20$, three levels} &
\includethreewide{\FastDtwentyFfive}{\BOSDtwentyFfive}{\HybridDtwentyFfive}{$M = 20$, five levels} \\
\includethreewide{\FastDtwentyFten}{\BOSDtwentyFten}{\HybridDtwentyFten}{$M = 20$, 10 levels} &
\includethreewide{\FastDtwentyFtwenty}{\BOSDtwentyFtwenty}{\HybridDtwentyFtwenty}{$M = 20$, 20 levels}
\end{tabular}
\end{figure}

\begin{figure}[!t]
\begin{tabular}{rr}
\includethreewide{\FastDtwentyfiveFone}{\BOSDtwentyfiveFone}{\HybridDtwentyfiveFone}{$M = 25$, one level} &
\includethreewide{\FastDtwentyfiveFtwo}{\BOSDtwentyfiveFtwo}{\HybridDtwentyfiveFtwo}{$M = 25$, two levels} \\
\includethreewide{\FastDtwentyfiveFthree}{\BOSDtwentyfiveFthree}{\HybridDtwentyfiveFthree}{$M = 25$, three levels} & 
\includethreewide{\FastDtwentyfiveFfive}{\BOSDtwentyfiveFfive}{\HybridDtwentyfiveFfive}{$M = 25$, five levels} \\
\includethreewide{\FastDtwentyfiveFten}{\BOSDtwentyfiveFten}{\HybridDtwentyfiveFten}{$M = 25$, 10 levels} &
\includethreewide{\FastDtwentyfiveFtwenty}{\BOSDtwentyfiveFtwenty}{\HybridDtwentyfiveFtwenty}{$M = 25$, 20 levels}
\end{tabular}
\end{figure}

\begin{figure}[!t]
\begin{tabular}{rr}
\includethreewide{\FastDthirtyFone}{\BOSDthirtyFone}{\HybridDthirtyFone}{$M = 30$, one level} &
\includethreewide{\FastDthirtyFtwo}{\BOSDthirtyFtwo}{\HybridDthirtyFtwo}{$M = 30$, two levels} \\
\includethreewide{\FastDthirtyFthree}{\BOSDthirtyFthree}{\HybridDthirtyFthree}{$M = 30$, three levels} &
\includethreewide{\FastDthirtyFfive}{\BOSDthirtyFfive}{\HybridDthirtyFfive}{$M = 30$, five levels} \\
\includethreewide{\FastDthirtyFten}{\BOSDthirtyFten}{\HybridDthirtyFten}{$M = 30$, 10 levels} &
\includethreewide{\FastDthirtyFtwenty}{\BOSDthirtyFtwenty}{\HybridDthirtyFtwenty}{$M = 30$, 20 levels} \\
\end{tabular}
\label{last-fig}
\end{figure}

\end{document}